\documentstyle[12pt]{article}
\begin{document}

\title{\Large\bf Reducible systems and embedding procedures in the
canonical formalism}

\author{\\
R. Banerjee\thanks{\noindent e-mail: rabin@if.ufrj.br --
Permanent address: S.N. Bose National Centre for Basic Sciences,
Block JD, Sector III, Salt Lake, Calcutta- 700091 , India.}~ and J.
Barcelos-Neto\thanks{\noindent 
e-mail: barcelos@if.ufrj.br}\\ 
Instituto de F\'{\i}sica\\
Universidade Federal do Rio de Janeiro\\ 
RJ 21945-970 - Caixa Postal 68528 - Brasil\\}
\date{}

\maketitle
\abstract
We propose a systematic method of dealing with the canonical
constrained structure of reducible systems in the Dirac and
symplectic approaches which involves an enlargement of phase and
configuration spaces, respectively. It is not necessary, as in the
Dirac approach, to isolate the independent subset of constraints or
to introduce, as in the symplectic analysis, a series of lagrange
multipliers-for-lagrange multipliers.  This analysis illuminates the
close connection between the Dirac and symplectic approaches of
treating reducible theories, which is otherwise lacking. The example
of $p$- form gauge fields ($p=2,3$) is analyzed in details.

\vfill\noindent
Total number of the manuscript pages: 24

\vspace{1cm}

\vspace{1cm}
\newpage

\section{Introduction} 
\renewcommand{\theequation}{1.\arabic{equation}}
\setcounter{equation}{0}

\bigskip
The problem of giving a proper formulation for reducible constrained
systems, be it in the canonical Hamiltonian\cite{D}, sumplectic
\cite{FJ}, or path integral \cite{HT} approaches, is quite involved
demanding a modification of the usual rules.  These systems occur
whenever the set of constraints found by the usual canonical
prescription is not linearly independent. Perhaps their most popular
occurrence is in the theory of p-form gauge fields involving
completely antisymmetric p-rank tensors, but these are also present
in other examples. Different proposals exist to account for the
reducibility property depending on the manifestation of the problem.
In the canonical approach, for instance, the Dirac brackets cannot be
computed since the corresponding matrix is singular and hence
noninvertible. A possible remedy is to isolate the independent subset
of constraints \cite{D,HT}. Then the Dirac brackets are computed
within this set following the normal procedure. The process is then
extended to include the complete set of constraints. This was the
approach adopted in \cite{K} for analyzing the 2-form gauge theory.
An alternative canonical method \cite{BN} based on the symplectic
form also suffers from an identical problem. As is known, the
constraints are obtained from the zero modes of the symplectic matrix
and inserted back in the Lagrangian through multipliers, in analogy
with the  usual Dirac method of introducing constraints in the
Hamiltonian. For reducible systems the symplectic matrix is
noninvertible.  It is cured by imposing additional conditions on the
Lagrange multipliers.  Therefore a series of Lagrange
multipliers-for-Lagrange multipliers is involved.  Details of this
approach can be found in \cite{BN}. In the BRST path integral
formulation, on the other hand, the functional measure is ill defined
because reducibility leads to presence of $\delta(0)$ terms. This is
usually avoided by introducing extra ghost fields.  Depending on the
degree of reducibility, a tower of ghosts-for-ghosts may be necessary
\cite{HT}.

\medskip
It is quite clear that the problem of reducibility manifests in
different ways leading to different suggestions for their treatment.
But several unanswered and unpleasant issues prevail. For instance,
there is no unique and systematic way of identifying the independent
subset of constraints in the Dirac approach. Additionally, such an
abstraction may lead to the loss of important symmetries of the
problem, as has been pointed out recently \cite{JK}. In both the
symplectic and BRST formalisms, on the contrary, it is not clear
whether the tower of extra fields is really necessary or an artifact
of the prescription. Furthermore, apparently there does not seem to
be any correlation among the different available resolutions of the
same problem. This is all the more surprising since the Dirac
\cite{D} and symplectic approaches \cite{FJ} are known to be
completely equivalent while the path integral can always be derived
from the canonical formalism.

\medskip
The preceding comments show that the problem of reducibility merits a
closer examination. This is the motivation of the present paper where
we attempt to provide answers to some of these basic issues. A
systematic canonical formalism for reducible systems in both the
Dirac and symplectic viewpoints  has been developed in details, while
the path integral has been left for a forthcoming work \cite{RB}. In
the Dirac approach it is shown that a suitable phase space extension,
involving a single pair of canonical fields, accounts for the
reducibility. This is true irrespective of the degree of reducibility.
It is not necessary to isolate an independent subset of constraints.
In the symplectic approach, on the contrary, the constraints are
embedded in an extended configuration space. Apart from the Lagrange
multipliers, which occur even for irreducible systems, there are two
extra fields which can be identified with the additional canonical
pair in the Dirac approach. This properly accounts for the
reducibility. The generalized brackets following from the symplectic
matrix agree with the Dirac brackets.

\medskip
Our ideas are introduced in a simple setting by discussing a quantum
mechanical toy model in section II. These  ideas are elaborated, in
section III, to cope with reducibility in Dirac's hamiltonian
formalism.  The examples of the 2-form and 3-form gauge field
theories are worked out in details. The analysis is next repeated in
section IV using the symplectic lagrangian formulation. Section V
contains our concluding remarks.

\vspace{1cm}
\section{A toy model} 
\renewcommand{\theequation}{2.\arabic{equation}}
\setcounter{equation}{0}

\bigskip
In this section a quantum mechanical model is considered to introduce
the ideas in a simple setting. The full power and utility of the
approach will be elaborated in the subsequent sections.

\medskip
Consider the following set of reducible constraints,

\begin{eqnarray}
T_a&=&p_a+\epsilon_{ab}\,q_b\approx0
\hspace{1.5cm} (\epsilon_{12}=1
\hspace{.5cm} a,b=1,2)
\nonumber\\
T_3&=&p_1+p_2-q_1+q_2\approx0
\nonumber\\
T_4&=&p_1-p_2+q_1+q_2\approx0
\label{2.1}
\end{eqnarray}

\bigskip\noindent
where ($q_a,p_a$) is a canonical set of variables.  It is clear that
only two of these constraints are independent. For convenience,
choose them to be $T_a$. Then the other constraints are expressed by
the combinations,

\begin{eqnarray}
T_3&=&T_1+T_2\approx0
\nonumber\\
T_4&=&T_1-T_2\approx0
\label{2.2}
\end{eqnarray}

\bigskip\noindent
It is simple to see that the usual Poisson brackets (PB) among the
canonical variables are incompatible with the above constraints. A
standard way to overcome this in the canonical formalism is to work
with the Dirac brackets (DB).  For computing these brackets it is
necessary to obtain the inverse of the matrix formed by the PB of the
complete set of constraints. In this case although the constraints
(\ref{2.1}) are second-class, the inverse does not exist because of
the reducibility condition (\ref{2.2}). The usual approach
\cite{D,HT} is to isolate the independent set of constraints and
evaluate the DB. These brackets will then strongly enforce
(\ref{2.1}). The matrix elements of the PB of the independent
constraints is given by,

\begin{equation}
S_{ab}=\bigl\{T_a,T_b\bigr\}
=2\,\epsilon_{ab}
\label{2.3}
\end{equation}

\bigskip\noindent
which has the following inverse 

\begin{equation}
S^{-1}_{ab}=-\,\frac{1}{2}\,\epsilon_{ab}
\label{2.4}
\end{equation}

\bigskip\noindent
Then the DB defined by the general formula \cite{D},

\begin{equation}
\bigl\{Q,P\bigr\}^\ast=\bigl\{Q,P\bigr\}
-\bigl\{Q,T_a\bigr\}S^{-1}_{ab}\bigl\{T_b,P\bigr\}
\label{2.5}
\end{equation}

\bigskip\noindent
are found to be,

\begin{eqnarray}
&&\bigl\{q_a,p_b\bigr\}^\ast=\frac{1}{2}\,\delta_{ab}
\nonumber\\
&&\bigl\{q_a,q_b\bigr\}^\ast=\bigl\{p_a,p_b\bigr\}^\ast
=-\,\frac{1}{2}\,\epsilon_{ab}
\label{2.6}
\end{eqnarray}

\bigskip\noindent
which strongly imposes the constraint sector (\ref{2.1}). This
completes the conventional treatment.

\medskip
In our approach, on the other hand, it is possible to work with the
full set of reducible constraints by first extending the phase space,
introducing a canonical pair of variables ($\eta,\pi$),

\begin{eqnarray}
&&\bigl\{\eta,\pi\bigr\}=1
\nonumber\\
&&\bigl\{\eta,\eta\bigr\}=\bigl\{\pi,\pi\bigr\}=0
\label{2.6a}
\end{eqnarray}

\bigskip\noindent
These variables have vanishing brackets with $q_a$ and $p_a$. In the
extended space the dependent constraints are modified as
\footnote{In those cases when the dependent constraints cannot be
easily separated, all the constraints have to be modified. Later on
we consider such examples.},

\begin{eqnarray}
\tilde T_3&=&p_1+p_2-q_1+q_2+2c\eta\approx0
\nonumber\\
\tilde T_4&=&p_1-p_2+q_1+q_2+c\pi\approx0
\label{2.7}
\end{eqnarray}

\bigskip\noindent
where $c$ is an arbitrary parameter and the factor 2 is included only
for computational ease. The matrix of the PB of the complete set of
constraints ($T_a,\tilde T_3,\tilde T_4$), which are now independent
in the extended space, is given by,

\begin{equation}
\tilde S=2\,
\left(\begin{array}{cccc}
0&1&1&-1\\
-1&0&-1&-1\\
-1&1&0&-2+c^2\\
1&1&2-c^2&0
\end{array}\right)
\label{2.8}
\end{equation}

\bigskip\noindent
The inverse is,

\begin{equation}
\tilde S^{-1}=\frac{1}{2c^2}\,
\left(\begin{array}{cccc}
0&2-c^2&-1&1\\
c^2-2&0&1&1\\
1&-1&0&-1\\
-1&-1&1&0
\end{array}\right)
\label{2.9}
\end{equation}

\bigskip\noindent
Expectedly, $\tilde S^{-1}$ does not exist for $c=0$. The DB are now
modified as,

\begin{equation}
\bigl\{Q,P\bigr\}^\ast=\bigl\{Q,P\bigr\}
-\bigl\{Q,\tilde T\bigr\}\tilde S^{-1}
\bigl\{\tilde T,P\bigr\}
\label{2.9a}
\end{equation}

\bigskip\noindent
where $\tilde T$ generically denotes the constraints ($T_a,\tilde
T_3,\tilde T_4$). A simple algebra reproduces (\ref{2.6}). It is
interesting to point out that the parameter $c$ is canceled in the
evaluation of these DB. This is related to the vanishing of $\eta$
and $\pi$ if $T_a$ are imposed in (\ref{2.7}). In other words, the
phase space extension removes the reducibility but retains the
original constraint sector independent of the value of $c$. This is
the reason that the DB of the reducible system were reproduced
without the need of taking any limit like $c\rightarrow0$ at the end
of the computations.

\vspace{1cm}
\section{The Dirac formalism} 
\renewcommand{\theequation}{3.\arabic{equation}}
\setcounter{equation}{0}

\bigskip
The Dirac formalism provides a systematic way of discussing the
canonical constrained structure of different systems. If such systems
are reducible, however, the usual analysis must be modified, as
already elaborated in the toy model. In this section we extend our
approach to the examples of 2-form and 3-form gauge field theories.
Two schemes will be developed, with and without a parameter in the
extended phase space.

\bigskip
\subsection{Two-form with a mass parameter}

\bigskip
The Lagrangian density is defined by,

\begin{equation}
{\cal L}=\frac{1}{12}\,H_{\mu\nu\rho}H^{\mu\nu\rho}
\label{3.1}
\end{equation}

\bigskip\noindent
where, 

\begin{equation}
H_{\mu\nu\rho}=\partial_\mu A_{\nu\rho}
+\partial_\rho A_{\mu\nu}+\partial_\nu A_{\rho\mu}
\label{3.2}
\end{equation}

\bigskip\noindent
is the field tensor corresponding to the 2 form gauge field
$A_{\mu\nu}$. 

\medskip
The canonical momenta are given by

\begin{eqnarray}
T_0&=&\pi_{00}\approx0
\label{3.3}\\
\pi_{ij}&=&\dot A_{ij}+\partial_iA_{j0}-\partial_jA_{i0}
\label{3.4}
\end{eqnarray}

\bigskip\noindent
from which the total Hamiltonian density is obtained

\begin{equation}
{\cal H}=\frac{1}{4}\,\pi_{ij}\pi^{ij}
+\partial_i A_{0j}\pi^{ij}
-\frac{1}{4}\,A_{ij}\nabla^2A^{ij}
+\frac{1}{2}\,A_{ij}\partial^j\partial_kA^{ik}
+\lambda_0\,T_0
\label{3.5}
\end{equation}

\bigskip\noindent
Persistence in time of the primary constraint $T_0\approx0$ leads to
a secondary constraint

\begin{equation}
T_i=\partial^j\pi_{ji}\approx0
\label{3.6}
\end{equation}

\bigskip\noindent

The constraints $T_i$ are reducible since,

\begin{equation}
\partial^iT_i=\partial^i\partial^j\pi_{ji}=0
\label{3.7}
\end{equation}

\bigskip\noindent
implies that all $T_i$ are not independent. This is related to the
fact that the original gauge transformations,

\begin{equation}
\delta A_{ij}=\partial_i\xi_j-\partial_j\xi_i
\label{3.8}
\end{equation}

\bigskip\noindent
are not independent since $\delta A_{ij}=0$ if the parameters are
$\xi_i=\partial_i\theta$ for any $\theta$. The conventional
gauge-fixing in the Dirac procedure is to choose \cite{HT,K},

\begin{equation}
\chi_i=\partial^jA_{ji}
\label{3.9}
\end{equation}

\bigskip\noindent
which also satisfies a reducibility condition like (\ref{3.7}). Due
to this condition,

\begin{eqnarray}
S_{ij}(\vec x,\vec y)&=&
\bigl\{T_i(\vec x),\chi_j(\vec y)\bigr\}
\nonumber\\
&=&-\,\bigl(\eta_{ij}\,\nabla^2
+\partial_i\partial_j\bigr)\,\delta(\vec x-\vec y)
\label{3.10}
\end{eqnarray}

\bigskip\noindent
does not possess an inverse
\footnote{We work with the Bjorken-Drell metric
$\eta_{ij}=-\delta_{ij}$.}.  
Hence the DB cannot be computed in the usual way. The orthodox method
\cite{D,HT,K} is to isolate the independent subset of constraints.
Apart from a lack of uniqueness in the procedure, the ensuing algebra
is quite messy~\cite{K}.

\medskip 
In the present approach we proceed, as before, by modifying the
constraints in an enlarged phase space,

\begin{eqnarray}
\tilde T_i&=&\partial^j\pi_{ji}+m\,p_i\approx0
\nonumber\\
\tilde\chi_i&=&\partial^jA_{ji}+m\,\phi_i\approx0
\label{3.11}
\end{eqnarray}

\bigskip\noindent
where ($\phi_i,p^j$) is a canonical pair,

\begin{equation}
\bigl\{\phi_i(\vec x),p^j(\vec y)\bigr\}
=\delta_i^j\,\delta(\vec x-\vec y)
\label{3.12}
\end{equation}

\bigskip\noindent
and $m$ is some parameter having the dimensions of mass. The matrix
involving the PB of the modified constraints is,

\begin{eqnarray}
\tilde S_{ij}(\vec x,\vec y)&=&\left(\begin{array}{cc}
\bigl\{\tilde T_i(\vec x),\tilde T_j(\vec y)\bigr\}
&\bigl\{\tilde T_i(\vec x),\tilde\chi_j(\vec y)\bigr\}\\
\bigl\{\tilde\chi_i(\vec x),\tilde T_j(\vec y)\bigr\}
&\bigl\{\tilde\chi_i(\vec x),\tilde\chi_j(\vec y)\bigr\}
\end{array}\right)
\nonumber\\
&&
\nonumber\\
&=&\left(\begin{array}{cc}
0&-1\\
1&0
\end{array}\right)
\Bigl[\bigl(\nabla^2+m^2\bigr)\,
\eta_{ij}+\partial_i\partial_j\Bigr]\,
\delta(\vec x-\vec y)
\label{3.13}
\end{eqnarray}

\bigskip\noindent
whose inverse is given by,

\begin{equation}
\tilde S_{ij}^{-1}(\vec x,\vec y)=\left(\begin{array}{cc}
0&1\\
-1&0
\end{array}\right)\,
\Bigl(\eta_{ij}-\frac{1}{m^2}\,
\partial_i\partial_j\Bigr)\,
\frac{1}{\nabla^2+m^2}\,
\delta(\vec x-\vec y)
\label{3.14}
\end{equation}

\bigskip\noindent
which will be used for computing the DB. These brackets are,

\begin{eqnarray}
&&\bigl\{A_{ij}(\vec x),\pi_{kl}(\vec y)\bigr\}^\ast
=\bigl\{A_{ij}(\vec x),\pi_{kl}(\vec y)\bigr\}
\nonumber\\
&&\phantom{\bigl\{A_{ij}(\vec x),}
-\int d\vec zd\vec w\,
\bigl\{A_{ij}(\vec x),\tilde T_n(\vec z)\bigr\}\,
\bigl(\tilde S^{-1}_{nr}(\vec z,\vec w)\bigr)_{12}
\bigl\{\tilde\chi_r(\vec w),\pi_{kl}(\vec y)\bigr\}
\nonumber\\
&&\phantom{\bigl\{A_{ij}(\vec x),}
=\Bigl(\eta_{ij,kl}
+\frac{\partial^2_{ij,kl}}{\nabla^2+m^2}\Bigr)\,
\delta(\vec x-\vec y)
\label{3.15}
\end{eqnarray}

\bigskip\noindent
where we have used some of the definitions below that shall be
considered throughout this paper

\begin{eqnarray}
\eta_{ij,kl}&=&\eta_{ik}\eta_{jl}-\eta_{il}\eta_{jk}
\label{3.16a}\\
\eta_{ijk,lmn}&=&\eta_{il}\eta_{jm}\eta_{kn}
+\eta_{in}\eta_{jl}\eta_{km}
+\eta_{im}\eta_{jn}\eta_{kl}
\nonumber\\
&&-\eta_{il}\eta_{jn}\eta_{km}
-\eta_{im}\eta_{jl}\eta_{kn}
-\eta_{in}\eta_{jm}\eta_{kl}
\label{3.16b}\\
\partial_{i,jk}&=&\eta_{ij}\partial_k-\eta_{ik}\partial_j
\label{3.16c}\\
\partial_{ij,klm}&=&\eta_{ij,kl}\partial_m
+\eta_{ij,mk}\partial_l+\eta_{ij,lm}\partial_k
\label{3.16d}\\
\partial_{ij,kl}^2&=&\eta_{ik}\partial_j\partial_l
+\eta_{jl}\partial_i\partial_k
-\eta_{jk}\partial_i\partial_l-\eta_{il}\partial_j\partial_k
\label{3.16e}
\end{eqnarray}

\bigskip\noindent
In the limit $m\rightarrow0$, this reproduces the standard result
\cite{K}. It is worthwhile to mention that the final result for the
DB is obtained only after taking the limit $m\rightarrow0$ because
this is essential for mapping (\ref{3.11}) to the original constraint
shell. Naturally, it would be conceptually cleaner if, as was done in
the toy model, the phase space extension accounts for the
reducibility but does not deform the original constraint sector. Then
the relevant DB will be obtained directly. This formulation is
presented now in the next section.

\bigskip
\subsection{Two-form without a mass parameter}

\bigskip
As an alternative approach which also illuminates the gauge-fixing in
the path integral quantization of reducible systems, the original
constraints are extended as

\begin{eqnarray}
\bar T_i&=&\partial^j\pi_{ji}+\partial_ip\approx0
\nonumber\\
\bar\chi_i&=&\partial^j A_{ji}+\partial_i\phi\approx0
\label{3.19}
\end{eqnarray}

\bigskip\noindent
where ($\phi,p$) denote a canonical set in the enlarged phase
space.  It is important to note that on the new constraint surface
$\partial^i\bar T_i=\partial^i\bar\chi_i\approx0$ imply
$\phi=p\approx0$, provided reasonable boundary conditions are
assumed. This phase space extension, therefore, simultaneously avoids
the reducibility and enforces the original constraints
$T_i=\chi_i\approx0$, irrespective of any limiting procedure, and the
correct DB of the original theory ought to be reproduced. This is
reminiscent of the quantum mechanical example. Incidentally, the
extended gauge condition $\bar\chi_i$ is precisely used for
gauge-fixing in the path integral BRST analysis
\cite{HT} where $\phi$ plays the role of the ghost field. By
eliminating the reducibility, terms like $\delta(0)$ no longer occur
in the path integral. A modified constraint like $\bar T_i$ was
suggested earlier \cite{HT1}, though in a different context.

\medskip
The new constraint matrix analogous to (\ref{3.13}) is now given by,

\begin{equation}
\bar S_{ij}(\vec x,\vec y)=
\eta_{ij}\,\left(\begin{array}{cc}
0&-1\\
1&0
\end{array}\right)\,
\nabla^2\,\delta(\vec x-\vec y)
\label{3.20}
\end{equation}

\bigskip\noindent
The inverse matrix is,

\begin{equation}
\bar S_{ij}^{-1}(\vec x,\vec y)
=\eta_{ij}\,\left(\begin{array}{cc}
0&1\\
-1&0
\end{array}\right)\,
\frac{1}{\nabla^2}\,\delta(\vec x-\vec y)
\label{3.21}
\end{equation}

\bigskip\noindent
The only nonvanishing DB is easily computed,

\begin{eqnarray}
\bigl\{A_{ij}(\vec x),\pi_{kl}(\vec y)\bigr\}^\ast
&=&\bigl\{A_{ij}(\vec x),\pi_{kl}(\vec y)\bigr\}
\nonumber\\
&&-\int d\vec zd\vec w\,
\bigl\{A_{ij}(\vec x),\bar T_n(\vec z)\bigr\}\,
\bigl(\bar S^{-1}_{nr}(\vec z,\vec w)\bigr)_{12}
\bigl\{\bar\chi_r(\vec w),\pi_{kl}(\vec y)\bigr\}
\nonumber\\
&=&\Bigl(\eta_{ij,kl}
+\frac{\partial^2_{ij,kl}}{\nabla^2}\Bigr)\,
\delta(\vec x-\vec y)
\label{3.22}
\end{eqnarray}

\bigskip\noindent
which reproduces the familiar expression. Both the elegance and
algebraic simplification in obtaining the result are noteworthy.
Therefore, from now on we shall only consider extensions analogous to
(\ref{3.19}) which avoid the necessity of introducing any parameter
and a subsequent limiting prescription.

\medskip
It is interesting to observe that the originally canonical pair
($\phi,p$) now has vanishing DB either among them or with the other
fields $A_{ij}$, $\pi_{ij}$. A similar phenomenon occurs in the
symplectic analysis discussed later.

\bigskip
\subsection{Three-form case}

\bigskip
The theory of a three form gauge field presents features that are
peculiar and representative of higher form examples. Consequently,
the analysis given here can be easily implemented to such examples.
The Lagrangian is now given by,

\begin{equation}
{\cal L}=\frac{1}{48}\,H_{\mu\nu\rho\lambda}H^{\mu\nu\rho\lambda}
\label{3.22a}
\end{equation}

\bigskip\noindent
where 

\begin{equation}
H_{\mu\nu\rho\lambda}=\partial_\mu A_{\nu\rho\lambda}
-\partial_\lambda A_{\mu\nu\rho}
+\partial_\rho A_{\lambda\mu\nu}
-\partial_\nu A_{\rho\lambda\mu}
\label{3.22b}
\end{equation}

\bigskip\noindent
is the fully antisymmetric field tensor written in terms of the
3-form field. By following the canonical Dirac procedure, the
reducible constraint is easily obtained,

\begin{equation}
T_{ij}=\partial^k\pi_{kij}\approx0
\label{3.23}
\end{equation}

\bigskip\noindent
where the momentum conjugate to $A^{ijk}$ is given by

\begin{equation}
\pi_{ijk}=\delta^{lmn}_{ijk}\,
\Bigl(\frac{1}{6}\,\dot A_{lmn}
-\frac{1}{2}\,\partial_lA_{mn0}\Bigr)
\label{3.23a}
\end{equation}

\bigskip\noindent
and we have used the notation given by Eq. (\ref{3.16b}).

\medskip
The corresponding Coulomb-like gauge fixing condition reads,

\begin{equation}
\chi_{ij}=\partial^kA_{kij}\approx0
\label{3.24}
\end{equation}

\bigskip\noindent
The PB matrix among these constraints, defined analogously to
(\ref{3.13}), is,

\begin{equation}
S_{ijkl}(\vec x,\vec y)=\left(\begin{array}{cc}
0&-1\\
1&0
\end{array}\right)
\Bigl(\eta_{ij,kl}\,\nabla^2
+\partial^2_{ij,kl}\Bigr)\,\delta(\vec x-\vec y)
\label{3.25}
\end{equation}

\bigskip\noindent
This quantity does not have an inverse which is defined as
\footnote{The factor 1/2 is necessary for avoiding double counting
due to the antisymmetry in the repeated indices $k,l$.}

\begin{equation}
\frac{1}{2}\int dy\,S_{ijkl}(x,y)S^{-1\,klmn}(y,z)
=\left(\begin{array}{cc}
1&0\\
0&1
\end{array}\right)\delta_{ij}^{mn}\,\delta(x-z)
\label{3.26}
\end{equation}

\bigskip
Exactly as was done for the two-form gauge field, it is possible to
extend the phase space such that the constraints are now modified as,

\begin{eqnarray}
\bar T_{ij}&=&\partial^k\pi_{kij}
+\partial_ip_j-\partial_jp_i\approx0
\nonumber\\
\bar\chi_{ij}&=&\partial^k A_{kij}
+\partial_i\phi_j-\partial_j\phi_i\approx0
\label{3.36}
\end{eqnarray}

\bigskip\noindent
where, as usual, the canonical set ($\phi_i,p^j$) in the enlarged
space has vanishing PB with the original variables. There is,
however, an important distinction from the analysis in the two form
example. In that case, the enlargement (\ref{3.19}) implied the vanishing
of the extra fields on the constraint surface thereby reproducing the
original constraint sector. Here, on the contrary, $\partial^i\bar
T_{ij}=\partial^i\bar\chi_{ij}\approx0$ leads to,

\begin{equation}
\bigl(\eta_{ij}\,\nabla^2+\partial_i\partial_j\bigr)\,p_j
=\bigl(\eta_{ij}\,\nabla^2+\partial_i\partial_j\bigr)\,\phi_j\approx0
\label{3.37}
\end{equation}

\bigskip\noindent
Clearly, it is not possible to set $p_i=\phi_i\approx0$ since these
are multiplied by a noninvertible operator. Hence by itself the
extension (\ref{3.36}) does not reduce to the original constraint
sector. There are two ways to overcome this situation. The phase
space is extended further by introducing more fields and performing a
fresh analysis. This is the analog of the BRST analysis where a
tower of ghosts-for-ghosts etc.  has to be inserted \cite{HT}.
Alternatively, the symplectic structure can be altered so that the
new fields are no longer canonical but satisfy,

\begin{equation}
\bigl\{\phi_i(\vec x),p_j(\vec y)\bigr\}
=\Bigl(\eta_{ij}+\frac{\partial_i\partial_j}{\nabla^2}\Bigr)\,
\delta(\vec x-\vec y)
\label{3.38}
\end{equation}

\bigskip\noindent
This means that these fields are transversal so that desired
inversion in (\ref{3.37}) is possible. The new fields vanish on the
constraint surface and the original constraint set is reproduced. A
related observation is that (\ref{3.38}) just corresponds to the DB
evaluated for the transversal constraints
$\partial_i\phi^i=\partial_ip^i\approx0$. It is now crucial to note
that the algebra of the antisymmetric combination of new fields in
(\ref{3.36}) is, however, unaffected by the deformation (\ref{3.38})
and yields the same result as if the fields were canonical. We find,

\begin{equation}
\bigl\{\partial_ip_j(\vec x)-\partial_jp_i(\vec x),
\partial_k\phi_l(\vec y)-\partial_l\phi_k(\vec y)\bigr\}
=\partial^2_{ij,kl}\,\delta(\vec x-\vec y)
\label{3.39}
\end{equation}

\bigskip\noindent
irrespective of whether the Poisson algebra or the modified
(\ref{3.38}) is used. This is understandable since (\ref{3.39})
involves the brackets among gauge invariant variables in which case
there should be no difference between the Poisson and modified
(Dirac) algebras \cite{HT}.  In the evaluation of PB among $\bar
T_{ij}$ and $\bar\chi_{ij}$, therefore, the new fields, which vanish
on the constraint surface, can still be chosen as canonical. The DB
of the original reducible theory will be obtained by working with the
modified constraints (\ref{3.36}) without the necessity of
introducing  either additional tower of fields, as in the BRST
approach, or deforming the canonical structure.

\medskip
The nontrivial PB among the constraints (\ref{3.36}) is

\begin{eqnarray}
\bigl(\bar S_{ijkl}(\vec x,\vec y)\bigr)_{12}
&=&\bigl\{\bar T_{ij}(\vec x),\bar\chi_{kl}(\vec y)\bigr\}
\nonumber\\
&=&-\,\eta_{ij,kl}\,\nabla^2\,\delta(\vec x-\vec y)
\label{3.40}
\end{eqnarray}

\bigskip\noindent
which has the following inverse,

\begin{equation}
\bigl[\bigl(\bar S_{ijkl}(x,y)\bigr)^{-1}\bigr]_{21}
=-\,\eta_{ij,kl}\,\frac{1}{\nabla^2}\,\delta(\vec x-\vec y)
\label{3.41}
\end{equation}

\bigskip\noindent
The complete inverse analogous to (\ref{3.21}) follows trivially. Now
the nontrivial brackets are easily computed,

\begin{equation}
\bigl\{A_{ijk}(\vec x),\pi_{lmn}(\vec y)\bigr\}^\ast
=\Bigl(\eta_{ijk,lmn}+\frac{\partial^2_{ijk,lmn}}{\nabla^2}\Bigr)\,
\delta(\vec x-\vec y)
\label{3.42}
\end{equation}

\bigskip\noindent
where the first term is the PB and the second is generated by
(\ref{3.41}). This yields the DB of the original reducible theory. It
is simple to generalize this approach to arbitrary $p$-form gauge
fields. All the arguments given above are applicable and the DB can
be evaluated by enlarging the constraint sector with a single
canonical pair of ($p-1$)-form field.

\vspace{1cm}
\section{The symplectic formalism} 
\renewcommand{\theequation}{4.\arabic{equation}}
\setcounter{equation}{0}

\bigskip
The symplectic formalism is a geometrical manner of dealing with
canonical systems. Although it existed in the contemporary literature
\cite{SM}, it was resurrected by Faddeev-Jackiw \cite{FJ} from a
physicist's point of view. Since the subject is still evolving
\cite{BN1} it is reasonable to provide a brief overview before
proceeding with the actual computations.

\medskip
The symplectic formalism deals with first-order Lagrangians. It is
opportune to mention that this is not a serious restriction because
all systems we know, described by quadratical Lagrangians, can always
be set in the first-order formulation. This is achieved by extending
the configuration space with the introduction of auxiliary fields.
For algebraic simplifications, these are usually the momenta, but
this is not necessary. The symplectic formalism is, therefore,
basically a Lagrangian approach complementing the Hamiltonian
formulation of Dirac.

\medskip
Let us consider a system described by a first-order Lagrangian such
as 

\begin{equation}
L=a_\alpha(y)\,\dot y^\alpha-V(y)
\label{4.1}
\end{equation}

\bigskip\noindent
where $y^\alpha$ is a set of $2N$ coordinates. The momenta or other
auxiliary quantities required to render the Lagrangian in the
first-order form will be denoted by $y^{i+N}$. We develop the
formalism by using discrete degrees of freedom. The extrapolation for
fields can be done in a straightforward way.

\medskip
From (\ref{4.1}), the Euler-Lagrange equation of motion reads

\begin{equation}
f_{\alpha\beta}\,\dot y^\beta=\partial_\alpha V
\label{4.2}
\end{equation}

\bigskip\noindent
where $\partial_\alpha\equiv\partial/\partial y^\alpha$ and

\begin{equation}
f_{\alpha\beta}=\partial_\alpha a_\beta-\partial_\beta a_\alpha
\label{4.3}
\end{equation}

\bigskip\noindent
is the symplectic 2-form. If $\det(f_{\alpha\beta})\not=0$, the
system is unconstrained and one can solve eq. (\ref{4.2}) for
velocities $\dot y^\alpha$, i.e., 

\begin{equation}
\dot y^\alpha=f^{\alpha\beta}\,\partial_\beta\,V
\label{4.4}
\end{equation}

\bigskip\noindent
where $f^{\alpha\beta}$ is the inverse of $f_{\alpha\beta}$. The
generalized brackets, which are the Poisson brackets, between the
coordinates $y^\alpha$ and $y^\beta$ are given by $f^{\alpha\beta}$,
i.e. 

\begin{equation}
\{y^\alpha,y^\beta\}=f^{\alpha\beta}
\label{4.4a}
\end{equation}

\bigskip
An interesting and instructive point occurs when the quantity
$f_{\alpha\beta}$ is singular. In this case one cannot identify the
symplectic tensor and, consequently, the brackets of the theory
cannot be consistently defined. This means that the system is
constrained from a symplectic point of view.
\footnote{We mention that the number of constraints in the symplectic
formalism is equal to or lesser than in the Dirac case. However, the
functional form is the same.}  
To identify and incorporate the constraints in this approach, we
proceed as discussed below.

\medskip
Let us denote the above mentioned singular quantity by
$f_{\alpha\beta}^{(0)}$, and suppose that it has, for example, $M$
$(M<2N)$ zero modes $v_m^{(0)}$, $m=1,\dots,M$, i.e.,

\begin{equation}
f_{\alpha\beta}^{(0)}\,v_m^{(0)\beta}=0
\label{4.5}
\end{equation}

\bigskip\noindent
The combination of (\ref{4.2}) and (\ref{4.5}) gives 

\begin{equation}
\tilde v_m^{(0)\alpha}\,\partial_\alpha V^{(0)}=0
\label{4.6}
\end{equation}

\bigskip\noindent
This may lead to a constraint. Let us suppose that this actually
occurs (we shall discuss the opposite case soon). Usually, as for
instance in the Dirac approach, the constraints are introduced in the
potential part of the Lagrangian by means of Lagrange multipliers.
Here, in order to get a deformation in the tensor
$f_{\alpha\beta}^{(0)}$ we introduce them instead into the kinetic
part. This is done by taking the time derivative of the constraint
and putting them in the Lagrangian by means of multipliers
\footnote{Since constraints do not evolve in time, a time derivative
of a constraint is also a constraint. Another point is that one
could, instead, take the time derivative of the Lagrange multiplier.
The difference, being a total derivative, does not affect the
equation of motion}.
These multipliers, which we denote by $\lambda_m^{(0)}$, enlarge the
configuration space of the theory. This permits us to identify new
vectors $a_\alpha^{(1)}$ and $a_m^{(1)}$ as

\begin{eqnarray}
a_\alpha^{(1)}&=&a_\alpha^{(0)}
+\lambda_m^{(0)}\,\partial_\alpha\Omega_m^{(0)}
\nonumber\\
a_m^{(1)}&=&0
\label{4.7}
\end{eqnarray}

\bigskip\noindent
where $\Omega_m^{(0)}$ are the constraints obtained from (\ref{4.6}).
Consequently, one can now introduce the quantities defining the
elements of the deformed symplectic matrix in the extended
configuration space ($y^\alpha,\lambda^{(0)}_m$),

\begin{eqnarray}
f_{\alpha\beta}^{(1)}&=&\partial_\alpha a_\beta^{(1)}
-\partial_\beta a_\alpha^{(1)}
\nonumber\\
f_{\alpha m}^{(1)}&=&\partial_\alpha a_m^{(1)}
-\partial_m a_\alpha^{(1)}
=-\partial_m a_\alpha^{(1)}
\nonumber\\
f_{mn}^{(1)}&=&\partial_m a_n^{(1)}-\partial_n a_m^{(1)}
\label{4.8}
\end{eqnarray}

\bigskip\noindent
where $\partial_m=\partial/\partial\lambda^m$. If det $f^{(1)}\neq0$,
then the process of finding the constraints terminates. If not, one
should repeat the above iterative procedure as many times as
necessary.

\medskip
It may also occur that we arrive at a stage where the zero modes of
the singular matrix do not lead to any new constraint. This is the
case, for example, of gauge theories. At this point, in order to
define the symplectic tensor, some gauge condition has to be imposed.
For details, see Ref. \cite{FJ}. For reducible systems, further
complications arise. The usual technique is to mimic the BRST
analysis \cite{HT} and introduce a series of Lagrange
multiplier-for-Lagrange multipliers \cite{BN}. In this section we
develop in details, both for two and three form gauge fields, the
analogous procedure already discussed in Dirac's formalism.

\bigskip
\subsection{The two-form case}

\bigskip
To use the symplectic formalism, it is necessary to express the
Lagrangian (\ref{3.1}) in the first order notation. For convenience,
this Lagrangian is rewritten as,

\begin{eqnarray}
{\cal L}&=&\frac{1}{4}\,\dot A_{ij}\dot A^{ij}
+\partial^iA^{j0}\,\dot A_{ij}
+\frac{1}{2}\,\partial_iA_{0j}\,\partial^iA^{0j}
+\frac{1}{4}\,\partial_iA_{jk}\,\partial^iA^{jk}
\nonumber\\
&&\phantom{\frac{1}{4}\,\dot A_{ij}\dot A^{ij}}
-\frac{1}{2}\,\partial_iA_{j0}\,\partial^jA^{i0}
+\frac{1}{2}\,\partial_iA_{jk}\,\partial^kA^{ji}
\label{4.9}
\end{eqnarray}

\bigskip\noindent
Using the momentum conjugate to $A_{ij}$ as the auxiliary field, we
can put (\ref{4.9}) in the desired first order form,

\begin{equation}
{\cal L}=-\,\frac{1}{4}\,\pi_{ij}\pi^{ij}
+\frac{1}{2}\,\Bigl(\dot A_{ij}
+\partial_iA_{j0}-\partial_jA_{i0}\Bigr)\,\pi^{ij}
+\frac{1}{4}\,\partial_iA_{jk}\,\partial^iA^{jk}
+\frac{1}{2}\,\partial_iA_{jk}\,\partial^kA^{ji}
\label{4.10}
\end{equation}

\bigskip\noindent
This is conveniently expressed as,

\begin{equation}
{\cal L}^{(0)}=\frac{1}{2}\,\pi_{ij}\,\dot A^{ij}-V^{(0)}
\label{4.11}
\end{equation}

\bigskip\noindent
where,

\begin{equation}
V^{(0)}=\frac{1}{4}\,\pi_{ij}\pi^{ij}
+\partial_iA_{0j}\pi^{ij}
-\frac{1}{4}\,A_{ij}\nabla^2A^{ij}
+\frac{1}{2}\,A_{ij}\,\partial^j\partial_kA^{ik}
\label{4.12}
\end{equation}

\bigskip\noindent
A comparison with the general structure (\ref{4.1}) leads to the
following identifications

\begin{eqnarray}
a_{ij}^{(0)A}&=&\pi_{ij}
\nonumber\\
a_i^{(0)A}&=&0
\nonumber\\
a_{ij}^{(0)\pi}&=&0
\label{4.13}
\end{eqnarray}

\bigskip\noindent
Nonvanishing elements of the symplectic matrix are therefore given
by, 

\begin{eqnarray}
f_{ijkl}^{(0)A\pi}(\vec x,\vec y)
&=&\frac{\delta a_{kl}^{(0)\pi}(\vec y)}{\delta A^{ij}(\vec x)}
-\frac{\delta a_{ij}^{(0)A}(\vec x)}{\delta\pi^{kl}(\vec y)}
\nonumber\\
&=&-\,\eta_{ij,kl}\,\delta(\vec x-\vec y)
\label{4.14}
\end{eqnarray}

\bigskip\noindent
Correspondingly, the matrix $f^{(0)}$, whose general form reads

\begin{equation}
f^{(0)}=\left(\begin{array}{ccc}
f_{ik}^{(0)AA}&f_{ikl}^{(0)AA}&f_{ikl}^{(0)A\pi}\\
f_{ijk}^{(0)AA}&f_{ijkl}^{(0)AA}&f_{ijkl}^{(0)A\pi}\\
f_{ijk}^{(0)\pi A}&f_{ijkl}^{(0)\pi A}&f_{ijkl}^{(0)\pi\pi}\\
\end{array}\right)
\label{4.15}
\end{equation}

\bigskip\noindent
can be written as,

\begin{equation}
f^{(0)}=\left(\begin{array}{ccc}
0&0&0\\
0&0&-\eta_{ij,kl}\\
0&\eta_{ij,kl}&0\\
\end{array}\right)\,\delta(\vec x-\vec y)
\label{4.16}
\end{equation}

\bigskip\noindent
This is clearly a singular matrix, which is exactly the way
constraints are manifested in the symplectic formalism. Let us
consider that a zero mode of (\ref{4.16}) has the general form:
$(v^k,u^{kl},\omega^{kl})$, where $u^{kl}$ and $\omega^{kl}$ are
antisymmetric quantities. Possible new constraints might appear from

\begin{equation}
\int d\vec x\biggl(v^k\frac{\delta}{\delta A^{0k}}
+\frac{1}{2}\,u^{kl}\frac{\delta}{\delta A^{kl}}
+\frac{1}{2}\,\omega^{kl}\frac{\delta}{\delta\pi^{kl}}\biggr)
\int d\vec y\,V^{(0)}=0
\label{4.17}
\end{equation}

\bigskip\noindent
For $(v^k,u^{kl},\omega^{kl})$ to be a zero mode of $f^{(0)}$, we
have 

\begin{eqnarray}
\eta_{ij,kl}\,\omega^{kl}=0
&\Longrightarrow&\omega_{ij}=0
\nonumber\\
\eta_{ij,kl}\,u^{kl}=0
&\Longrightarrow&u_{ij}=0
\label{4.18}
\end{eqnarray}

\bigskip\noindent
and the quantity $v^k$ remains undetermined. Using these results and
the expression for $V^{(0)}$ given in (\ref{4.12}), it is found that
(\ref{4.17}) yields,

\begin{equation}
\int d\vec x\,v^k(\vec x)\,
\frac{\delta}{\delta A^{0k}(\vec x)}\int d\vec y\,V^{(0)}
=\int d\vec x\,v_j\partial_i\pi^{ij}=0
\label{4.19}
\end{equation}

\bigskip\noindent
Since $v_j$ is a generic function of $\vec x$, we obtain the
constraint 

\begin{equation}
T^j=\partial_i\pi^{ij}
\label{4.20}
\end{equation}

\bigskip\noindent
This is how the secondary constraint of the Dirac formalism manifests
in the symplectic version. We mention that primary constraints in the
Dirac approach are not constraints in the symplectic case. Now,
proceeding to the next step in the iterative process,

\begin{eqnarray}
{\cal L}^{(1)}&=&\frac{1}{2}\,\pi_{ij}\dot A^{ij}
+\lambda_j\,\bigl(\partial_i\dot\pi^{ij}+\partial^j\dot p\bigr)
-V^{(0)}
\nonumber\\
&=&\frac{1}{2}\,\pi_{ij}\dot A^{ij}
-\partial_i\lambda_j\,\dot\pi^{ij}
-\partial^j\lambda_j\dot p-V^{(1)}
\label{4.21}
\end{eqnarray}

\bigskip\noindent
where

\begin{equation}
V^{(1)}=\frac{1}{4}\,\pi_{ij}\pi^{ij}
-\frac{1}{4}\,A_{ij}\nabla^2A^{ij}
+\frac{1}{2}\,A_{ij}\,\partial^j\partial_kA^{ik}
\label{4.22}
\end{equation}

\bigskip\noindent
The term $\partial_iA_{0j}\pi^{ij}$ in $V^{(0)}$ was absorbed by the
term $\partial_i\lambda_j\dot\pi^{ij}$. Hence, the $A_{0i}$ field has
disappeared from the theory. Note also that the reducible constraint
(\ref{4.20}) has been modified following the same pattern as in
(\ref{3.19}). The new coefficients are

\begin{eqnarray}
a_{ij}^{(1)A}&=&\pi_{ij}
\nonumber\\
a_{ij}^{(1)\pi}&=&-\partial_i\lambda_j+\partial_j\lambda_i
\nonumber\\
a_i^{(1)\lambda}&=&0
\nonumber\\
a^{(1)p}&=&-\,\partial_i\lambda^i
\label{4.23}
\end{eqnarray}

\bigskip\noindent
The matrix $f^{(1)}$, introduced in (\ref{4.8}), now has the general
form 

\begin{equation}
f^{(1)}=\left(\begin{array}{cccc}
f^{(1)AA}_{ijkl}&f^{(1)A\pi}_{ijkl}
&f^{(1)A\lambda}_{ijk}&f^{(1)Ap}_{ij}\\
&&&\\
f^{(1)\pi A}_{ijkl}&f^{(1)\pi\pi}_{ijkl}
&f^{(1)\pi\lambda}_{ijk}&f^{(1)\pi p}_{ij}\\
&&&\\
f^{(1)\lambda A}_{ikl}&f^{(1)\lambda\pi}_{ikl}
&f^{(1)\lambda\lambda}_{ik}&f^{(1)\lambda p}_i\\
&&&\\
f^{(1)pA}_{kl}&f^{(1)p\pi}_{kl}
&f^{(1)p\lambda}_k&f^{(1)pp}\\
\end{array}\right)
\label{4.25}
\end{equation}

\bigskip\noindent
where the notation follows (\ref{4.14}), except that the coefficients
$a^{(0)}$ are replaced by $a^{(1)}$. A simple algebra leads to the
result, 
\begin{equation}
f^{(1)}=\left(\begin{array}{cccc}
0&-\eta_{ij,kl}&0&0\\
\eta_{ij,kl}&0&\partial_{k,ij}&0\\
0&\partial_{i,lk}&0&\partial_i\\
0&0&\partial_k&0\\
\end{array}\right)\,\delta(\vec x-\vec y)
\label{4.26}
\end{equation}

\bigskip\noindent
This matrix is still singular so that the iterative process has to be
continued.

\medskip
Let us consider that the zero mode of (\ref{4.26}) is given by
($v^{kl},u^{kl},\omega^k,h$), where $v^{kl}$ and $u^{kl}$ are
antisymmetric quantities. This implies

\begin{eqnarray}
&&\eta_{ij,kl}\,u^{kl}=0
\hspace{.5cm}\Longrightarrow u_{ij}=0
\nonumber\\
&&\eta_{ij,kl}\,v^{kl}+\partial_{k,ij}\,\omega^k=0
\nonumber\\
&&\phantom{(\eta_{ik}\eta_{jl}}
\Longrightarrow v_{ij}=\frac{1}{2}\,
\Bigl(\partial_i\omega_j-\partial_j\omega_i\Bigr)
\nonumber\\
&&\partial_i h=0
\nonumber\\
&&\partial_k\omega^k=0
\label{4.27}
\end{eqnarray}

\bigskip\noindent
In order to look for new constraints, we write

\begin{equation}
\int d\vec x\,\biggl(\frac{1}{2}\,v^{kl}\,
\frac{\delta}{\delta A^{kl}}
+\omega^k\,\frac{\delta}{\delta\lambda^k}
+h\,\frac{\delta}{\delta\pi}\biggr)
\int d\vec y\,V^{(1)}=0
\label{4.28}
\end{equation}

\bigskip\noindent
The l.h.s. of this equation vanishes identically. Thus, the zero mode
of $f^{(1)}$ does not lead to a new constraint. This fact means that
we are in the presence of a gauge theory. Let us then choose the
Coulomb-like gauge condition already considered in Eq. (\ref{3.19}),

\begin{equation}
\partial_iA^{ij}+\partial^j\phi=0
\label{4.30}
\end{equation}

\bigskip\noindent
Introducing this constraint into the kinetic part of the Lagrangian,
we have

\begin{eqnarray}
{\cal L}^{(2)}&=&\frac{1}{2}\,\pi_{ij}\,\dot A^{ij}
-\partial_i\lambda_j\,\dot\pi^{ij}
-\partial_i\lambda^i\,\dot p
+\eta_j(\partial_i\dot A^{ij}+\partial^j\dot\phi)
-V^{(1)}
\nonumber\\
&=&\frac{1}{2}\,\Bigl(\pi_{ij}
-\partial_i\eta_j+\partial_j\eta_i\Bigr)\,\dot A^{ij}
-\partial_i\lambda_j\dot\pi^{ij}
-\partial_i\lambda^i\dot\pi
-\partial_i\eta^i\dot\phi-V^{(2)}
\label{4.31}
\end{eqnarray}

\bigskip\noindent
where 

\begin{equation}
V^{(2)}=\frac{1}{4}\,\pi_{ij}\pi^{ij}
-\frac{1}{4}\,A_{ij}\nabla^2A^{ij}
\label{4.32}
\end{equation}

\bigskip\noindent
The term $\frac{1}{2}A_{ij}\partial^j\partial_kA^{ik}$ was absorbed
by $\partial_i\eta_j\dot A^{ij}$. The new coefficients are

\begin{eqnarray}
a^{(2)A}_{ij}&=&\pi_{ij}
-\partial_i\eta_j+\partial_j\eta_i
\nonumber\\
a^{(2)\pi}_{ij}&=&-\partial_i\lambda_j+\partial_j\lambda_i
\nonumber\\
a^{(2)p}&=&-\,\partial_i\lambda^i
\nonumber\\
a^{(2)\phi}&=&-\,\partial_i\eta^i
\nonumber\\
a^{(2)\lambda}_i&=&0
\nonumber\\
a^{(2)\eta}_i&=&0
\label{4.33}
\end{eqnarray}

\bigskip
The second iterated matrix $f^{(2)}$ is now calculated from
$a^{(2)}$, just as $f^{(1)}$ was done from $a^{(1)}$. We find 

\begin{equation}
f^{(2)}=\left(\begin{array}{cccccc}
0&-\eta_{ij,kl}&0&0&\partial_{k,ij}&0\\
\eta_{ij,kl}&0&\partial_{k,ij}&0&0&0\\
0&\partial_{i,lk}&0
&\partial_i&0&0\\
0&0&\partial_k&0&0&0\\
\partial_{i,lk}&0&0&0&0&\partial_i\\
0&0&0&0&\partial_k&0\\
\end{array}\right)\,\delta(\vec x-\vec y)
\label{4.35}
\end{equation}

\bigskip\noindent
where rows and columns follow the order $A_{ij}$, $\pi_{ij}$,
$\lambda_i$, $\pi$, $\eta_i$, and $\phi$. The above matrix is not
singular. Hence, it can be identified as the symplectic tensor of the
constrained theory. Its inverse will gives us the generalized
brackets of the physical fields of the theory.  The calculation of
the inverse is done in Appendix A. We simply write the final
result for the symplectic matrix

\begin{eqnarray}
f^{(2)^{-1}}&=&\left(\begin{array}{cccccc}
0&\delta_{ij}^{mn}+\frac{\partial_{ij}^{2\,mn}}{\nabla^2}&0&0
&\frac{\partial_i^{m,j}}{\nabla^2}&0\\
-\delta_{ij}^{mn}-\frac{\partial_{ij}^{2\,mn}}{\nabla^2}&0
&\frac{\partial_i^{m,j}}{\nabla^2}&0&0&0\\
0&\frac{\partial_{i,}^{nm}}{\nabla^2}&0
&-\frac{\partial^m}{\nabla^2}
&-\frac{1}{\nabla^2}\Bigl(\delta_i^m
+\frac{\partial_i\partial^m}{\nabla^2}\Bigr)&0\\
0&0&-\frac{\partial_i}{\nabla^2}&0&0&0\\
\frac{\partial_{i,}^{nm}}{\nabla^2}&0
&\frac{1}{\nabla^2}\Bigl(\delta_i^m
+\frac{\partial_i\partial^m}{\nabla^2}\Bigr)&0&0&
-\frac{\partial^m}{\nabla^2}\\
0&0&0&0&-\frac{\partial_i}{\nabla^2}&0\\
\end{array}\right)
\nonumber\\
&&\hspace{3cm}\times\delta(\vec x-\vec y)
\label{4.36}
\end{eqnarray}

\bigskip\noindent
Recalling the ordering of the fields, it is easy to read-off the
bracket between $A_{ij}$ and $\pi_{mn}$ from the (12) element of the
above matrix. This reproduces the DB given in (\ref{3.22}). Likewise,
other brackets are easily obtained.

\medskip
Incidentally, the symplectic brackets between the set ($\phi,p$)
reproduce the vanishing algebra found earlier in the Dirac formalism.
This illuminates the close connection between the embedding
procedures adopted in these two approaches.

\subsection{The three-form case}

\bigskip
The initial step in the symplectic formalism is to rewrite the
Lagrangian (\ref{3.22a}) in its first order form. It is convenient to
choose the momenta conjugate to $A_{ijk}$ as the auxiliary variables
in analogy with the 2-form example. We find

\begin{equation}
{\cal L}=\frac{1}{6}\,\pi_{ijk}\dot A^{ijk}-V^{(0)}
\label{4.37}
\end{equation}

\bigskip\noindent
where

\begin{eqnarray}
V^{(0)}&=&\frac{1}{12}\,\pi_{ijk}\pi^{ijk}
+\frac{1}{2}\,\partial_iA_{jk0}\pi^{ijk}
-\frac{1}{12}\,A_{ijk}\nabla^2A^{ijk}
\nonumber\\
&&\phantom{\frac{1}{12}\,\pi_{ijk}\pi^{ijk}}
+\frac{1}{4}\,\partial_iA^{ikl}\partial^jA_{jkl}
\label{4.38}
\end{eqnarray}

\bigskip\noindent
Following a similar procedure as in the previous subsection, we find the
constraint (\ref{3.23}). Adopting the same logic, this constraint is
modified by introducing new fields. Its structure is now identical to
the first relation in (\ref{3.36}.) At this point, there is an
important difference from the Dirac analysis. The transversality
condition on the new field,

\begin{equation}
\partial_ip^i=0
\label{4.39}
\end{equation}

\bigskip\noindent
is explicitly taken as an additional constraint.

\medskip
The corresponding gauge condition is given by the second relation in
(\ref{3.36}), together with a condition akin to (\ref{4.39}),

\begin{equation}
\partial_i\phi^i=0
\label{4.40}
\end{equation}

\bigskip
All these constraints are now introduced into the kinetic part of
the Lagrangian by means of Lagrange multipliers. We thus have

\begin{eqnarray}
{\cal L}&=&\frac{1}{6}\,\pi_{ijk}\dot A^{ijk}
+\frac{1}{2}\,\lambda_{jk}\,
\Bigl(\partial_i\dot\pi^{ijk}
+\partial^j\dot p^k-\partial^k\dot p^j\Bigr)
\nonumber\\
&&\phantom{\frac{1}{6}\,\pi_{ijk}\dot A^{ijk}}
+\frac{1}{2}\,\eta_{jk}\,
\Bigl(\partial_i\dot A^{ijk}
+\partial^j\dot\phi^k-\partial^k\dot\phi^j\Bigr)
\nonumber\\
&&\phantom{\frac{1}{6}\,\pi_{ijk}\dot A^{ijk}}
+\rho\,\partial_i\dot p^i
+\zeta\,\partial_i\dot\phi^i-V^{(0)}
\nonumber\\
&=&\frac{1}{6}\,\Bigl(\pi_{ijk}
-\partial_i\eta_{jk}-\partial_j\eta_{ki}
-\partial_k\eta_{ij}\Bigr)\,\dot A^{ijk}
-\frac{1}{6}\,\Bigl(\partial_i\lambda_{jk}+\partial_j\lambda_{ki}
+\partial_k\lambda_{ij}\Bigr)\,\dot\pi^{ijk}
\nonumber\\
&&\phantom{\frac{1}{6}\,\pi_{ijk}\dot A^{ijk}}
-\,(\partial_i\rho+\partial^j\lambda_{ji})\,\dot p^i
-\,(\partial_i\zeta+\partial^j\eta_{ji})\,\dot\phi^i-V^{(2)}
\label{4.41}
\end{eqnarray}

\bigskip\noindent
where

\begin{equation}
V^{(2)}=\frac{1}{12}\,\pi_{ijk}\pi^{ijk}
-\frac{1}{12}\,A_{ijk}\nabla^2A^{ijk}
\label{4.42}
\end{equation}

\bigskip
The symplectic coefficients are easily identified as,

\begin{eqnarray}
a_{ijk}^{(2)A}&=&\pi_{ijk}-\partial_i\eta_{jk}
-\partial_j\eta_{ki}-\partial_k\eta_{ij}
\nonumber\\
a_{ijk}^{(2)\pi}&=&-\partial_i\lambda_{jk}
-\partial_j\lambda_{ki}-\partial_k\lambda_{ij}
\nonumber\\
a_i^{(2)p}&=&-\partial_i\rho-\partial^j\lambda_{ji}
\nonumber\\
a_i^{(2)\phi}&=&-\partial_i\zeta-\partial^j\eta_{ji}
\nonumber\\
a_i^{(2)\rho}&=&0
\nonumber\\
a_i^{(2)\zeta}&=&0
\nonumber\\
a_i^{(2)\lambda}&=&0
\nonumber\\
a_i^{(2)\eta}&=&0
\label{4.42a}
\end{eqnarray}

\bigskip\noindent
The nonvanishing elements of the matrix $f^{(2)}$ are now computed.
Two elements are explicitly furnished to clarify the definitions and
notations.

\begin{eqnarray}
f^{(2)A\pi}_{ijklmn}(\vec x,\vec y)
&=&\frac{\delta a^{(2)\pi}_{lmn}(\vec y)}{\delta A^{ijk}(\vec x)}
-\frac{\delta a^{(2)A}_{ijk}(\vec x)}{\delta\pi^{lmn}(\vec y)}
\nonumber\\
&=&-\,\eta_{ijk,lmn}\,\delta(\vec x-\vec y)
\nonumber\\
f^{(2)A\eta}_{ijklm}(\vec x,\vec y)
&=&\frac{\delta a^{(2)\eta}_{lm}(\vec y)}{\delta A^{ijk}(\vec x)}
-\frac{\delta a^{(2)A}_{ijk}(\vec x)}{\delta\eta^{lm}(\vec y)}
\nonumber\\
&=&(\eta_{jk,lm}\partial_i+\eta_{ki,lm}\partial_j
+\eta_{ij,lm}\partial_k)\,\delta(\vec x-\vec y)
\label{4.43}
\end{eqnarray}

\bigskip\noindent
Likewise, all the entries are obtained to yield

\begin{eqnarray}
f^{(2)}&=&\left(\begin{array}{cccccccc}
0&-\eta_{ijk,lmn}&0&\partial_{lm,ijk}&0&0&0&0\\
\eta_{ijk,lmn}&0&\partial_{lm,ijk}&0&0&0&0&0\\
0&\partial_{ij,lmn}&0&0&\eta_{kl,ij}\partial^k&0&0&0\\
\partial_{ij,lmn}&0&0&0&0&\eta_{kl,ij}\partial^k&0&0\\
0&0&\eta_{ji,lm}\partial^j&0&0&0&\partial_i&0\\
0&0&0&\eta_{ji,lm}\partial^j&0&0&0&\partial_i\\
0&0&0&0&\partial_l&0&0&0\\
0&0&0&0&0&\partial_l&0&0\\
\end{array}\right)
\nonumber\\
&&\hspace{3cm}\times\delta(\vec x-\vec y)
\label{4.44}
\end{eqnarray}

\bigskip\noindent
where the matrix is arrayed in the sequence followed in (\ref{4.42a}).
Proceeding as in the 2-form theory, the inverse of this matrix is
calculated. From the appropriate entry in this symplectic matrix the
nontrivial bracket between $A_{ijk}$ and $\pi_{lmn}$ is easily
obtained. It reproduces the algebra (\ref{3.42}) found in the Dirac
method.

\vspace{1cm}
\section{Conclusion}

\bigskip
A general scheme for treating reducible systems has been developed in
the canonical formalism, both from the Hamiltonian (Dirac) and
Lagrangian (symplectic) viewpoints. This avoids either the
abstraction of the independent subset of constraints, or the
introduction of an infinite set of new fields - manipulations that
are essential in the conventional analysis of reducible theories
\cite{D,HT,K,BN}. Apart from being systematic, a significant feature
was the algebraic simplicity of the method. In this connection it may
be recalled that the usual Dirac method \cite{D,K} of obtaining the
algebra (\ref{3.22}) appears quite involved compared to this
calculation. 

\medskip
Within the Dirac approach, the present scheme consisted in a phase
space extension involving only a single canonical pair. There are two
interesting aspects of this procedure. The first is that the present
phase space extension achieves exactly the opposite of a similar
approach \cite{BFFT} designed to convert second class constraints
into their first class forms. The point is that the second class
reducible constraints discussed here yield a vanishing determinant
for the constraint Poisson bracket matrix. Roughly speaking,
therefore, these constraints continue to display a first class
character. The phase space embedding changes this character into true
second class, enabling a simple and direct evaluation of the Dirac
brackets. The second aspect is that by imposing restrictions on the
single extra canonical pair of variables, the need for more fields
was avoided. This was explicitly demonstrated for the three form
gauge theory.

\medskip
The symplectic approach, in contrast to the Dirac procedure, is a
Lagrangian formulation. Nevertheless, the reducible constraints were
identified and then modified as in the Dirac treatment. The
generalized brackets obtained from the symplectic matrix agreed with
the Dirac brackets. An important observation concerns the distinct
roles played by the extra fields in the two approaches. In the Dirac
approach these fields, to begin with, formed a canonical pair.
However, at the end, the same fields were found to have vanishing
Dirac brackets either among them or with the other fields. In the
symplectic case, the new fields were just some multipliers, which
were obviously not canonical pairs since there is only an extension
of the configuration space. However, the symplectic matrix revealed
that these multipliers had vanishing brackets. Thus, although the
extra fields have different interpretations, they are algebraically
equivalent. This also shows, in a precise fashion, that the same
basic principle works for dealing with reducibility either in the
Dirac or symplectic formalism.

\medskip
A final question remains regarding the application of these ideas to
provide a path integral formulation for reducible systems. Some hints
can already be obtained from the present analysis. Recall that the
usual symplectic formalism requires a tower of Lagrange multipliers
\cite{BN} in analogy with a series of ghosts-for-ghosts necessary in
the BRST path integral formulation \cite{HT}. Since the tower of
Lagrange multipliers was avoided in this work, it suggests the
possibility that the corresponding situation in the BRST framework is
also redundant. Indeed, we shall explicitly show in a future
publication \cite{RB} that the path integral following from the
present canonical prescription eliminates the reducibility without
the necessity of any ghosts-for-ghosts.

\vspace{1cm}
\noindent
{\bf Acknowledgment:} This work is supported in part by Conselho
Nacional de Desenvolvimento Cient\'{\i}fico e Tecnol\'ogico - CNPq,
Financiadora de Estudos e Projetos - FINEP and Funda\c{c}\~ao
Universit\'aria Jos\'e Bonif\'acio - FUJB (Brazilian Research
Agencies). One of the authors (RB) would like to thank the members of
the Department of Theoretical Physics, UFRJ, for their kind
hospitality. 

\vspace{1cm}
\appendix
\renewcommand{\theequation}{A.\arabic{equation}}
\setcounter{equation}{0}
\section*{Appendix A}

\bigskip
Here we briefly outline the computation of the inverse of the matrix
$f^{(2)}$ given by (\ref{4.33}). We write the general form of the
inverse, which will give the symplectic matrix, as

\begin{equation}
f^{(2)^{-1}}=\left(\begin{array}{cccccc}
(AA)^{klmn}&(A\pi)^{klmn}&(A\lambda)^{klm}
&(Ap)^{kl}&(A\eta)^{klm}&(A\phi)^{kl}\\
(\pi A)^{klmn}&(\pi\pi)^{klmn}&(\pi\lambda)^{klm}
&(\pi p)^{kl}&(\pi\eta)^{klm}&(\pi\phi)^{kl}\\
(\lambda A)^{kmn}&(\lambda\pi)^{kmn}&(\lambda\lambda)^{km}
&(\lambda p)^k&(\lambda\eta)^{km}&(\lambda\phi)^k\\
(pA)^{mn}&(p\pi)^{mn}&(p\lambda)^m
&(pp)&(p\eta)^m&(p\phi)\\
(\eta A)^{kmn}&(\eta\pi)^{kmn}&(\eta\lambda)^{km}
&(\eta p)^k&(\eta\eta)^{km}&(\eta\phi)^k\\
(\phi A)^{mn}&(\phi\pi)^{mn}&(\phi\lambda)^m
&(\phi p)&(\phi\eta)^m&(\phi\phi)\\
\end{array}\right)
\label{A.1}
\end{equation}

\bigskip\noindent
This inverse is defined in such a way that

\begin{equation}
\int d\vec y\,f^{(2)}(\vec x,\vec y)\,
f^{(2)^{-1}}(\vec y,\vec z)
=\left(\begin{array}{cccccc}
\delta_{ij}^{mn}&0&0&0&0&0\\
0&\delta_{ij}^{mn}&0&0&0&0\\
0&0&\delta_i^m&0&0&0\\
0&0&0&1&0&0\\
0&0&0&0&\delta_i^m&0\\
0&0&0&0&0&1\\
\end{array}\right)\delta(\vec x-\vec z)
\label{A.2}
\end{equation}

\bigskip\noindent
By a straightforward matrix algebra a set of equations is obtained, the
solutions of which yield the desired entries in (\ref{A.1}). The
calculations are simplified by noting that the symplectic matrix must
posses the same symmetry structure as $f^{(2)}$. The final result is
explicitly displayed in (\ref{4.36}).

\end{document}